\documentclass[aps,showpacs,twocolumn]{revtex4}
\usepackage[cp1251]{inputenc}
\usepackage[russian]{babel}
\usepackage{graphicx}
\begin{document}

\title{\large \bf
Comment on \\ "Conventional superconductivity at 203 kelvin at
high pressures in \\ the sulfur hydride system"
 (A. P. Drozdov
et al., Nature 525, 73 (2015))}
\author{L.S.Mazov}
\affiliation{Institute of Physics of Microstructures, Russian Academy of Sciences Nizhny
Novgorod 603600 Russia}
\begin{abstract}
It is demonstrated that resistive transition at 203 K observed in
metallic sulfur hydride system at high pressure \cite{Droz1} can
be magnetic (rather than superconducting (SC)) in nature. The
onset temperature of genuine superconducting transition in these
compounds appears to be essentially lower on temperature. The
normal-state magnetic (AF SDW) phase transition preceding a
superconducting one ($T_c < T_m$) is characteristic for HTSC
cuprates, pnictides (selenides) and organic superconductors. The
resistive drop is provided by disappearing of magnetic (AF spin
fluctuation) scattering of conduction electrons and hence
formation of AF SDW order in the normal state. The formation of
such modulated magnetic structure in sulfur hydride seems to be
possible because of magnetic properties of metallic hydrogen at
high densities (in analogy with iron). Such unconventional picture
with two successive phase transitions: magnetic (AF SDW) and only
then superconducting one is naturally described by Keldysh-Kopaev
theory of dielectric (metal-insulator) phase transition in systems
with coexistence of superconducting (e-e) and dielectric (e-h)
pairings.
\end{abstract}
\maketitle

In a recent literature, there was claimed about significant
progress towards room temperature superconductivity in sulfur
hydrides at high pressure ("conventional superconductivity at 203
K" \cite{Droz1}, see also \cite{Droz2}). The conclusion about
superconducting nature of resistive transition at 203 K was made
on the basis of both sharp drop of resistivity below 203 K up to
nearly zero and shift of so-defined "onset temperature" towards
zero with magnetic field. The decrease of paramagnetic
magnetization beginning at the same temperature ($\sim 203 K$) was
considered in \cite{Droz1} as supporting fact for superconducting
nature of resistive transition. The noticeable shift of resistive
transition due to isotope effect under replacing of deuterium
instead of hydrogen permits them conclude about conventional
(phonon) nature of so-defined superconducting mechanism. However,
such formal attributes being necessary are not sufficient for
resistive transition to be superconducting.

As known, the thermodynamics of Keldysh-Kopaev (KK) dielectric
(metal-insulator) phase transition \cite{KK64} is the same as that
for superconductor. And in systems with coexistence of dielectric
(e-h) and superconducting (e-e) pairings superconducting
transition is preceded by dielectric phase transition ($T_c \le
T_D$). In such case, the dielectric gap $\Sigma$ is first opened
at symmetrical parts of the Fermi surface, and only at lower
temperature the superconducting gap $\Delta$ opens at the rest
part of the Fermi surface. In result, below critical temperature
of superconducting transition $T_c$ two different in nature gaps
coexist with one other in electron energy spectrum. The KK
dielectric transition provides some rise of critical temperature
of SC transition due to increase of electron density of states
(DOS) at the edges of dielectric gap due to removing of electronic
states from the energy region of dielectric gap. Such mechanism
for rise of $T_c$ remains to be phonon (but not conventional!) in
nature \cite{PHTSC}.
\begin{figure}[t]
\includegraphics[width=6.0 cm]{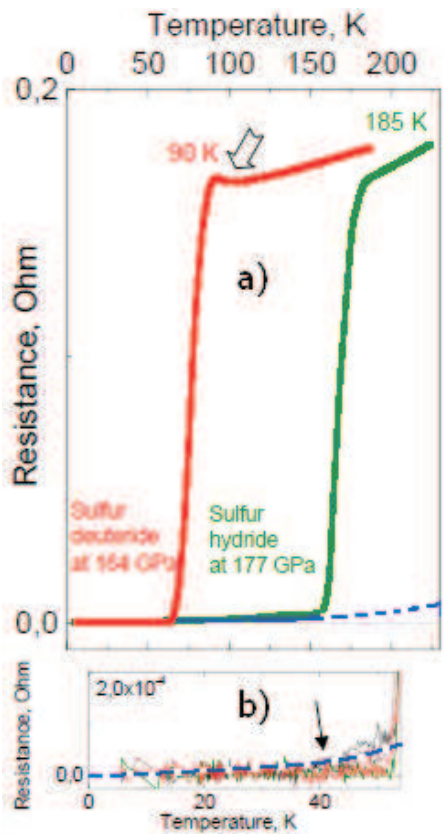}
\caption{The resistive transition curves for sulfur hydride (green
curve) and sulfur deuteride (red line) at high pressure. (a) high
temperature region: arrow indicates upturn part of resistive curve
characteristic for insulating behavior; blue dashed curve
schematically indicates the low-temperature part of BG (phonon)
curve (for details, see text); b) low temperature region: blue
dashed curve schematically demonstrates the intersection of BG
(phonon) curve with tail of total resistive transition,
determining the onset temperature of genuine SC transition
$T_c^{onset}$ (the data are from
\cite{Droz1,Droz2}).}\label{disp1}
\end{figure}
The sequence of two successive phase transitions: dielectric and
only then superconducting ($T_c < T_D$) is in fact realized in
high-$T_c$ cuprates \cite{M1,M2}, and organic superconductors
(see, e.g. \cite{Jer12}). The dielectric phase transition in these
systems is realized as magnetic (AF spin density wave (SDW),
incommensurate with lattice period) phase transition in the
conducting planes of doped compounds (normal state ($T_c < T_m =
T^*$)). Though this transition begins well in the normal state (at
$T \sim T^*$) but, due to intensive AF spin fluctuations, it is
completed only at lower temperature, when AF spin-fluctuation
scattering of conduction electrons disappears and modulated
magnetic structure appears with quite sharp resistive drop and
following then  by superconducting transition. These two
transitions are separated at total resistive transition curve by
well known Bloch-Gruneisen (BG) curve, characteristic for phonon
scattering in most metals \cite{Zim}. The resistivity above
BG-curve (normal state) is determined by scattering of conduction
electrons via AF spin-fluctuations, while below BG-curve the
phonons determine behavior of the system (including SC).

In such approach, the situation in sulfur hydrides, which
compounds become to be metallic at such high pressure, seems to be
quite similar to that in optimally-doped cuprates, pnictides and
organic superconductors (Fig.1). The sharp drop of resistivity at
203 K can be attributed to magnetic (AF SDW) phase transition when
AF spin fluctuation scattering of conduction electrons disappears
and the resistivity is determined only by scattering with phonons.
The indication to formation of AF SDW in the normal state is more
clear seen from resistivity measurements for deuterium hydride at
high pressure (red curve in Fig.1a) when sharp resistive
transition is preceded by upturn in resistive dependence, which
behavior is also characteristic for cuprates and pnictides
(selenides) in underdoped case. As for magnetic (AF spin
fluctuation) scattering of conduction electrons in sulfur hydrides
then it can be provided due to magnetic properties of metallic
hydrogen at such high pressure (see, e.g. \cite{Min84,Hir89}) . In
this sense, because of ferromagnetism in metallic hydrogen
\cite{Hir89}, the situation seems to be closer to Fe-based HTSC.
If so, then it can be proposed that hydrogen atoms in sulfur
hydride system under high pressure are arranged
antiferromagnetically (AF) leading to formation of AF SDW (cf.
with \cite{Min84}).

Since the Debye temperature $\Theta_D$ in these compounds is high
enough (of the order of 5000 K in our estimations) the phonon part
of resistivity at $T \le 203 K$ is relatively low (see schematic
blue dashed curves in Fig.1 a,b) and point of its intersection
with tail of total resistivity at lower temperature (corresponding
to the genuine onset temperature $T_c^{onset}$) is difficult to
determine correctly because of high enough noise level in
resistivity measurements at low-temperature (see, Fig.1b).
(Similar problem with correct determination of onset temperature
of SC transition appears in cuprates and pnictides at isothermal
resistivity measurements in high magnetic fields, see e.g.
\cite{M2}). From Fig.1 the onset temperature of genuine SC
transition can be estimated as near $T_c^{onset} \sim 40 K$. Such
estimation is consistent with observation of "a step with $T_c
\sim$ 30 K" noted in supplementary information to \cite{Droz2}. Of
course, to be more correct, it is necessary additional, more
detailed study of transport and other properties of these
compounds.

In other words, the above brief analysis of available resistive
data for sulfur hydrides at high pressure demonstrates that in
these compounds, at $T \sim 203 K$, a sharp resistive transition,
corresponding to the normal-state magnetic (AF SDW) phase
transition, occurs, and only then the system can enter the SC
state ($T_c < T_m$). The onset of the genuine, unconventional
superconducting transition in H$_2$S seems to be located only
noticeably lower on temperature ($T_c^{onset} \sim 40 K$).


\begin{thebibliography}{99}

\bibitem{Droz1} Drozdov A.P., Eremets M.I., Troyan I.A., Ksenofontov V.,
Shylin S.I. {\it Nature} {\bf 525} 73 (2015).

\bibitem{Droz2}  Drozdov A.P., Eremets M.I., Troyan I.A. {\it arXiv:cond-mat.}
{\bf 1412.0460} (2014).

\bibitem{KK64} Keldysh L.V., Kopaev Yu.V. {\it Sov.Phys.-Solid State} {\bf 6} 2792 (1964.)

\bibitem {PHTSC} {\it High Temperature Superconductivity} (Eds. V L Ginzburg and D A
    Kirzhnitz, Nauka Press, Moscow, 1977) [NY: Consultants Bureau,
    1982].

\bibitem{M1} Mazov L.S. {\it J.Low Temp.Phys.} {\bf 17}, 738 (1991);
{\it Phys.Rev. B} {\bf 70}, 054501 (2004).

\bibitem{M2} Mazov L.S. {\it Bull. of Russian Acad.of Sci.} {\bf 78}, 1348 (2014).

\bibitem {Jer12} Jerome D. {\it J.Supercond. Nov.Magn.} {\bf 25}, 633 (2012).

\bibitem{Zim} Ziman J.M. Electrons and Phonons (Oxford Univ.Press,
Oxford). 1960.

\bibitem{Min84} Min B.I., Jansen H.J.F., Freeman A.J. {\it Phys.Rev. B} {\bf 30} 5076
(1984).

\bibitem{Hir89} Hirsch J.E. {\it Phys.Lett. A} {\bf 141} 191
(1989).








\end{thebibliography}
\end{document}